\documentclass[a4paper,fleqn,usenatbib]{mnras}

\usepackage{newtxtext,newtxmath}
\usepackage[T1]{fontenc}
\usepackage{ae,aecompl}

\usepackage{graphicx} 
\usepackage{amsmath}  
\usepackage{amssymb}  
\usepackage{algpseudocode}
\usepackage{color}
\usepackage{enumitem}

\newcommand{\AU}{{\rm au}}
\newcommand{\Mj}{M_{\rm J}}
\newcommand{\Mo}{M_\odot}
\newcommand{\Ms}{M_\star}
\newcommand{\Me}{M_\oplus}

\title[Giant planet IMF]{Toward an initial mass function for giant planets}

\author[Carrera et al.]{
Daniel Carrera,\thanks{E-mail: danielc@astro.lu.se}
Melvyn B. Davies,
and Anders Johansen
\\
Lund Observatory, Department of Astronomy and Theoretical Physics, Lund University, Box 43, SE-221 00 Lund, Sweden\\
}

\date{Accepted XXX. Received YYY; in original form ZZZ}

\pubyear{2016}

\begin{document}
\label{firstpage}
\pagerange{\pageref{firstpage}--\pageref{lastpage}}
\maketitle

%
%
\begin{abstract}
%
%
The distribution of exoplanet masses is not primordial. After the initial stage of planet formation is complete, the gravitational interactions between planets can lead to the physical collision of two planets, or the ejection of one or more planets from the system. When this occurs, the remaining planets are typically left in more eccentric orbits.
%
%
Here we use present-day eccentricities of the observed exoplanet population to reconstruct the initial mass function of exoplanets before the onset of dynamical instability.
%
%
We developed a Bayesian framework that combines data from N-body simulations with present-day observations to compute a probability distribution for the planets that were ejected or collided in the past. Integrating across the exoplanet population, we obtained an estimate of the initial mass function of exoplanets.
%
%
We find that the ejected planets are primarily sub-Saturn type planets. While the present-day distribution appears to be bimodal, with peaks around $\sim 1 \Mj$ and $\sim 20 \Me$, this bimodality does not seem to be primordial. Instead, planets around $\sim 60 \Me$ appear to be preferentially removed by dynamical instabilities.
%
%
Attempts to reproduce exoplanet populations using population synthesis codes should be mindful of the fact that the present population has been depleted of intermediate-mass planets.
%
%
Future work should explore how the system architecture and multiplicity might alter our results.
\end{abstract}


%
%

%
%
\section{Introduction}
\label{sec:intro}

The past twenty years of exoplanet observations have revealed a large diversity of planetary systems \citep[e.g.][]{Mayor_2011,Borucki_2011,Batalha_2013}, including a large number of planets with orbital eccentricities much higher than those found in the solar system. These eccentricities challenged our understanding of planet formation occurring in a disk, where eccentricities are dampened by planet-disk interactions. Several authors have suggested that these high eccentricities might have a dynamical origin \citep[e.g.][]{Ford_2001,Juric_2008}. In this view, the eccentricities arise from planet-planet interactions which can cause a planetary system to become dynamically unstable. During a dynamical instability, close encounters between planets lead to either collisions or planet-planet scatterings. This dynamically active period ends only with a collision, or the ejection of one of the planets from the system. \citet{Juric_2008} showed that this type of process can reproduce the observed distribution of exoplanet eccentricities (at least those above $e = 0.2$) for a wide range of possible initial conditions. A similar result was obtained by \citet{Raymond_2009b}; these authors also showed how the eccentricities can be subsequently dampened by a planetesimal belt for less massive planets ($m_1 + m_2 < M_J$).

Several authors have noticed a mass-eccentricity correlation among exoplanets: low-mass planets tend to be on less eccentric orbits than high-mass planets \citep[e.g.][]{Ribas_2007,Ford_2008,Raymond_2010}. \citet{Raymond_2010} proposed that planet masses in the same system are correlated. Specifically, systems that form high-mass planets also tend to form equal-mass planets, while low-mass planet systems produce a wide diversity of mass ratios. The present work is an extension of this idea. We use the present-day eccentricities of observed exoplanets to estimate the masses of the planets that were ejected or collided. The result of this process is the exoplanet initial mass function (IMF).

This paper is organized as follows. In Section \ref{sec:stability} we summarize the theory behind dynamical stability and discuss previous work. In Section \ref{sec:bayesian} we present an introduction to Approximate Bayesian Computation, and describe how the algorithm can be adapted to the problem we want to solve. We describe our core simulations in Section \ref{sec:methods}, including our data selection method. The results of these simulations are presented in Section \ref{sec:results}. In Section \ref{sec:discussion} we discuss possible caveats and suggest opportunities for future research. Finally, we summarize and conclude in Section \ref{sec:conclusions}.

%
%
\section{Dynamical instability}
\label{sec:stability}

In this section we cover some important background and key results regarding the stability of planetary systems. In principle, the eccentricity of the surviving planet is determined by the need to conserve energy and angular momentum. For example, in the simple case of two planets, on coplanar orbits the energy $E$ and angular momentum $\Lambda$ of each planet is,

\begin{eqnarray}
	E &=& - \frac{G M m}{2a} \\
    \Lambda &=& M m \sqrt{\frac{G a (1 - e^2)}{M + m}},
\end{eqnarray}
where $M$ is the stellar mass, and $m$, $a$, and $e$ are the mass, semimajor axis, and eccentricity of the planet \citep[e.g.][]{Davies_2014}. In practice, this has limited utility because the problem is degenerate, even in the simplest case of a two-planet system. After a dynamical instability, the escaping planet carries an uncertain amount of angular momentum \citep{Ford_2001}. Even if the angular momentum of the ejected planet was known, one would still have only two equations for five unknowns (the mass of the ejected planet, and the initial mass and semimajor axis of the two planets). In sections \ref{sec:bayesian} and \ref{sec:methods} we describe the statistical approach that we use to tackle this degeneracy.

There are two common definitions of dynamical stability. In Hill stability, the ordering of the planets, in terms of their proximity to the central star, is fixed. In other words, the planet orbits never cross, but the outermost planet is allowed to escape the system. The other, more stringent definition, is Lagrange stability. In it, planets retain their ordering (i.e. are Hill stable), remain bound to the star, and variations in semimajor axis and eccentricity remain bounded. While Lagrange stability is the more useful definition, it has proven more difficult to solve mathematically than Hill stability.

There is no analytic solution for the evolution of three gravitating bodies, but in some cases it is possible to derive analytic constraints on the motion of the planets \citep[e.g.][]{Zare_1977}. These constraints can be understood as limitations on angular momentum exchange between planets. For example, \citet{Gladman_1993} showed that, to first order, a pair of planets on coplanar orbits are guaranteed to be Hill stable whenever

\begin{equation}\label{eqn:stability}
	\alpha^{-3} \left( \mu_1 + \frac{\mu_2}{\delta^2} \right)
    (\mu_1 \gamma_1 + \mu_2 \gamma_2 \delta)^2
    > 1 + 3^{4/3} \frac{\mu_1 \mu_2}{\alpha^{4/3}}
\end{equation}
where

\begin{eqnarray}
	\delta &=& \sqrt{\frac{a_2}{a_1}}\\
    \gamma_k &=& \sqrt{1 - e_k^2} \\
    \mu_k  &=& \frac{m_k}{M} \\
    \alpha &=& \mu_1 + \mu_2
\end{eqnarray}
and where $M$ is the mass of the star, $m_k$ is the mass of a planet, and $a$ and $e$ are the planet semimajor axis and eccentricity in barycentric coordinates. Therefore, for given planet masses and eccentricities, there is a critical value of $\delta$ that guarantees Hill stability \citep[see also the discussion in ][]{Barnes_2006}.

\citet{Raymond_2009} refer to the entire left-hand-side of Equation (\ref{eqn:stability}) as $\beta$, and the right-hand-side as $\beta_{\rm crit}$. They showed that planet-planet scatterings naturally lead to planetary systems at the edge of dynamical stability, with $\beta/\beta_{\rm crit}$ just above 1, and that this result is in agreement with observation. This is further evidence that dynamical instabilities driven by planet-planet scatterings occur frequently in planetary systems.

For a system of two low-mass planets with near-circular coplanar orbits, the Hill stability criterion (Equation (\ref{eqn:stability})) is approximately $\Delta > 2 \sqrt{3}$ \citep{Chambers_1996}, where $\Delta$ is the semimajor axis separation measured in mutual Hill radii,

\begin{eqnarray}\label{eqn:Delta}
	\Delta &=& \frac{a_2 - a_1}{R_H}, \\
	R_H   &=& \left( \frac{m_1 + m_2}{3 M} \right)^{1/3} \left( \frac{a_1 + a_2}{2} \right),
\end{eqnarray}
where $m_1$, $m_2$, $a_1$, and $a_2$ are the masses and semimajor axes of the two planets. For systems with more than two planets there is no analytic solution, but \citet{Chambers_1996} showed that these systems are probably unstable for separations up to $\Delta = 10$, at least for equal-mass planets. The time before planets experience their first close encounter ($t_{\rm ce}$) grows exponentially with $\Delta$. For a given value of $\Delta$, $t_{\rm ce}$ seems to depend weakly on the number of planets (for systems with more than two planets) and the planet masses \citep{Chambers_1996,Faber_2007}. For a planetary system with ten equal-mass planets, \citet{Faber_2007} found

\begin{equation}
	\log_{10} \left( \frac{t_{\rm ce}}{\rm yr} \right)
    \sim 	-5 - \log_{10} \left( \frac{\mu}{10^{-3}} \right)
    		+ 1.44 \Delta \left( \frac{\mu}{10^{-3}} \right)^{\frac{1}{12}},
\end{equation}
where $\mu = m/M$ is the planet-star mass ratio. This is an extremely steep dependence on $\Delta$, meaning that two systems with similar $\Delta$ values can have very different lifetimes.

%
%
\section{Approximate Bayesian Computation}
\label{sec:bayesian}

We saw in section \ref{sec:stability} that it is not possible to exactly determine the mass of an ejected exoplanet given the present-day observables. In this section we explain how one can use modern statistical methods together with N-body simulations to obtain a \textbf{probability distribution} for the ejected planet mass. Our tool of choice is Approximate Bayesian Computation, or ABC \citep[see the introduction by][]{Marin_2011}. Like all Bayesian algorithms, ABC is a way to solve the problem

\begin{equation}
	P(\theta|D) \; \propto \; P(D|\theta) P(\theta),
\end{equation}
where $D$ is some data set, $\theta$ is a set of model parameters, and $P(\theta)$ is the Bayesian prior. In its simplest form, the ABC algorithm looks like this

\begin{algorithmic}
\For{$i = 1:N$}
	\Repeat
	    \State $\theta' \gets P(\theta)$
	    \State $D' \gets {\rm model}(\theta')$
    \Until $\rho(D,D') < \epsilon$
    \State $\theta_i \gets \theta'$
\EndFor
\end{algorithmic}

The function $\rho$ is a distance measure that compares the synthetic dataset $D'$ and the observed data $D$, and $\epsilon$ is some critical value. The resulting collection of values $\{\theta_i\}$ approximately follows the distribution $P(\theta|D)$.

\subsection{Application: two-planet instability}
\label{sec:bayesian:application}

The simplest problem is a two-planet instability that always results in an ejection (i.e. planets are not allowed to collide). In this case, the data $D = (m, e)$ is the mass and final eccentricity of the observed planet, and the model parameter $\theta = q$ is the mass ratio of the ejected and remaining planets ($q = m_{\rm ej} / m_{\rm rem}$). We choose a uniform prior for $P(q)$ (we justify this choice in \ref{sec:methods:prior_q}). The ABC algorithm becomes

\begin{algorithmic}
\For{$i = 1:N$}
	\Repeat
	    \State $q' \gets P(q)$
	    \State $e' \gets {\rm simulation}(q')$
    \Until $|e - e'| < \epsilon$
    \State $q_i \gets q'$
\EndFor
\end{algorithmic}

In this investigation we choose $N = 1000$. Here we have overlooked the fact that the result of the N-body simulation depends on the initial separation between the two planets. Our approach is to choose a range of orbital separations that are consistent with planet formation: we select only the runs that remain stable for at least 0.5 Myr (see section \ref{sec:methods:p2}). Algorithmically, we can write this as

\begin{algorithmic}
\For{$i = 1:N$}
	\Repeat
	    \State $q' \gets P(q)$
	    \State $\Delta' \gets P(\Delta)$
	    \State $t', e' \gets {\rm simulation}(q', \Delta')$
    \Until $(|e - e'| < \epsilon) \;\; \& \;\; (t' > 0.5 \;{\rm Myr})$
    \State $q_i \gets q'$
\EndFor
\end{algorithmic}
where $t'$ is the time to the ejection. The final step is to allow collisions between planets and record whether or not a collision occurred. Let $C$ be a boolean value that is true when the simulation ends in a collision. The final algorithm becomes

\begin{algorithmic}
\For{$i = 1:N$}
	\Repeat
	    \State $q' \gets P(q)$
	    \State $\Delta' \gets P(\Delta)$
	    \State $C', t', e' \gets {\rm simulation}(q', \Delta')$
    \Until $(|e - e'| < \epsilon) \;\; \& \;\; (t' > 0.5 \;{\rm Myr})$
    \State $q_i \gets q'$
    \State $C_i \gets C'$
\EndFor
\end{algorithmic}
Given the list of $\{(q_i, C_i)\}$ one can construct a probability distribution of the masses of the two planets:

\begin{algorithmic}
\For{$i = 1:N$}
    \If{$C_i'$}
    	\State $m_{1,i} \gets m / (1+q_i')$
	\Else
    	\State $m_{1,i} \gets m$
    \EndIf
    \State $m_{2,i} \gets q_i * m_{1,i}$
\EndFor
\end{algorithmic}

The set of values $\{m_{1,i}, m_{2,i}\}$ approximately follow the posterior distribution $P(m_1, m_2| m, e)$. Finally, we repeat this process for every observed exoplanet that has a measured mass, and for each exoplanet we obtain a set $\{m_{1,i}, m_{2,i}\}$. Because all the sets have the same size ($N = 1000$), it is straight forward to combine them together. The result is an estimate of the exoplanet IMF.

In section \ref{sec:methods} we explain our simulations in greater detail, and we give a rationale for choosing a uniform prior for $P(q)$. We also explain how the stability requirement is implemented in practice.

%
%
\section{Our simulations}
\label{sec:methods}

In this section we describe how we selected our list of exoplanets, and we show that exoplanet observations support the use of a uniform prior $P(q)$ as described in section \ref{sec:bayesian}. We also explain our simulations in detail, and explain how exoplanet observations inform our choices of initial conditions. At the end of the section we explain how the stability criterion is implemented in detail.

\subsection{Data selection}
\label{sec:methods:data}

\begin{figure}
	\includegraphics[width=\columnwidth]{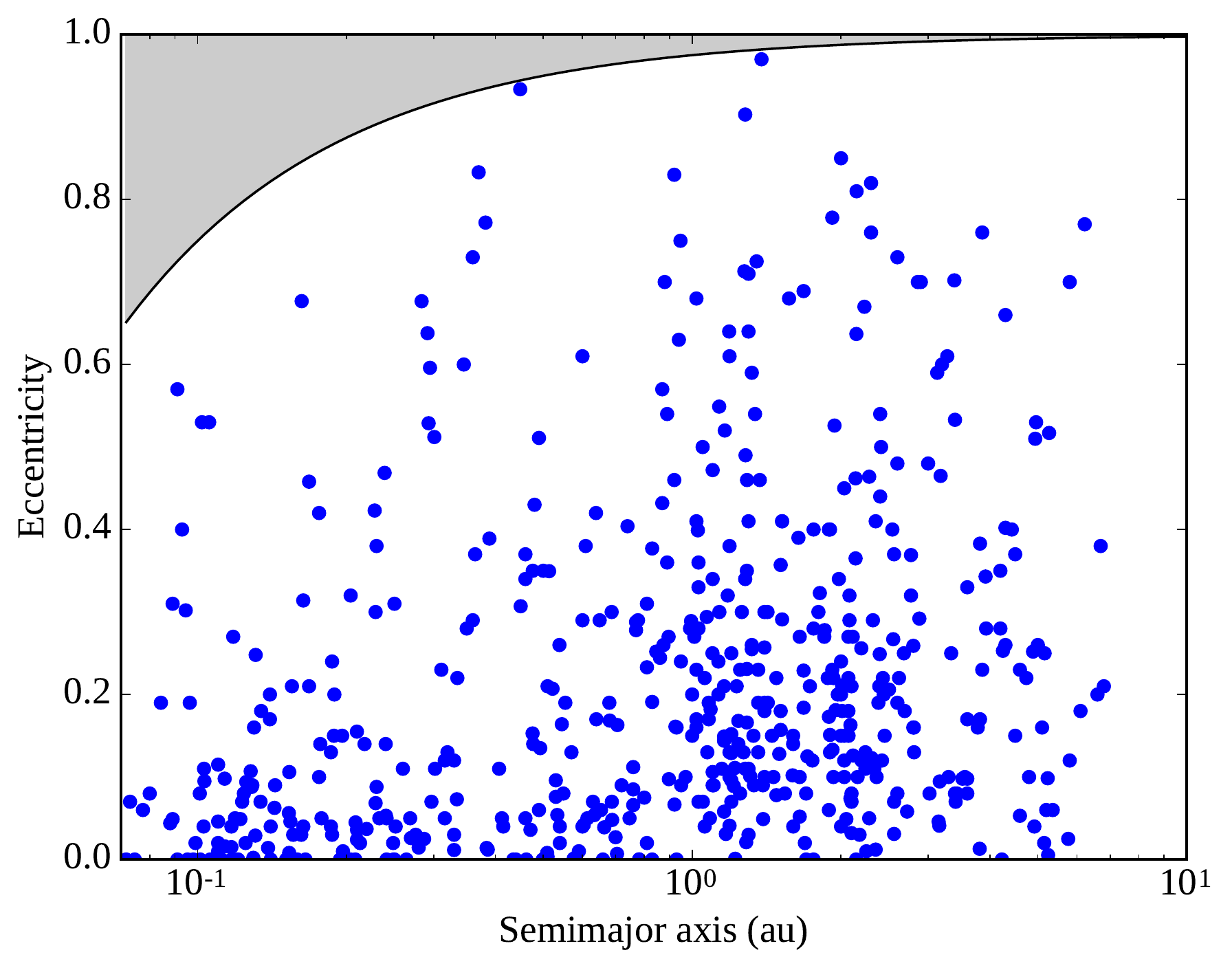}
    \caption{Orbital elements for all planets in our dataset. The grey region is a ``forbidden'' region where the the planet's periapsis is close enough to the star to cause tidal circularization \citep{Anderson_2016}. The planet data was obtained from the \texttt{exoplanet.eu} database on 16 Aug 2016. Table \ref{tab:dataset} has more information about our data set, and section \ref{sec:methods:data} describes our data selection.}
    \label{fig:data}
\end{figure}

We took the exoplanet catalogue from \texttt{exoplanets.eu} on 16 August 2016. We selected the planets that have mass measurements (i.e. we did not use any mass-radius relation). This means that our sample is dominated by planets discovered by the Radial Velocity method, meaning that we only have the planet's \textit{minimum mass}. We then exclude planets with orbital periods shorter than 10 days as many of those may have experienced high-eccentricity migration and tidal circularization. Our resulting dataset included six pulsar planets, so we removed those as well. This left us with 553 exoplanets. Figure \ref{fig:data} shows the semimajor axes and eccentricities of these planets; Table \ref{tab:dataset} also gives an overview of the dataset.

In this investigation we limit the dynamical simulations to giant planets. To define ``giant exoplanet'' we use the planet-star mass ratio $m / \Ms$, which we normalize with the Jupiter-Sun mass ratio for convenience

\begin{equation}\label{eqn:mu}
	\mu = \frac{m / \Ms}{\Mj/\Mo},
\end{equation}
where $\Mj$ is the mass of Jupiter. We define ``giant exoplanet'' as those with $0.05 < \mu < 10$. As a point of reference, Saturn has $\mu = 0.30$, Neptune has $\mu = 0.054$, and a super-Earth with $m = 10M_\oplus$ orbiting the Sun would have $\mu = 0.031$. Table \ref{tab:other-planets} lists the seven planets that were discovered by a method other than radial velocity or transit.

\begin{table}
	\centering
	\caption{An overview of the exoplanet dataset used in this work. In this work we assume that planets with $\mu < 0.05$ or $e < 0.05$ acquired their eccentricity without scattering another planet. Our data selection is described in section \ref{sec:methods:data}.}
	\label{tab:dataset}
	\begin{tabular}{llrr}
    	& & $\mu > 0.05$ & $\mu \le 0.05$ \\
		\hline
        \multicolumn{2}{l}{Total number of exoplanets}        & 506 & 47 \\
        \multicolumn{2}{l}{Eccentric ($e \ge 0.05$)}          & 427 & 15 \\
        \multicolumn{2}{l}{Discovered by radial velocity}     & 461 & 23 \\
        \multicolumn{2}{l}{Discovered by transit}             &  38 & 23 \\
        \multicolumn{2}{l}{Discovered by other methods}       &   7 &  1 \\
        \multicolumn{2}{l}{Members of multiple-giant systems} & 171 & 37 \\
        Around O stars   & ($\Ms/\Mo > 16$)                   &   0 &  0 \\
        Around AB stars  & ($1.5 < \Ms/\Mo \le 16$)           &  92 &  0 \\
        Around FGK stars & ($0.5 \le \Ms/\Mo \le 1.5$)        & 388 & 39 \\
        Around M stars   & ($\Ms/\Mo < 0.5$)                  &  26 &  8 \\
		\hline
	\end{tabular} \\
    \textbf{Note:} Mass limits from \citet{Habets_1981}.
\end{table}

\begin{table*}
	\centering
	\caption{List of planets included in our sample that were not discovered by either radial velocity or the transit method. The first planet has $\mu < 0.05$.}
	\label{tab:other-planets}
	\begin{tabular}{lrrrrrrl}
	Planet & mass ($\Mj$) &  period (day) & semimajor axis (AU) & eccentricity & star mass ($\Mo$) & Discovery method \\
		\hline
        KOI-620.02       & 0.024 &  130.19 & 0.51   & 0.008  & 1.0   & TTV          \\
		51 Eri b         & 7.0   & 14965.0 & 14.0   & 0.21   & 1.75  & Imaging      \\
		HD 176051 b      & 1.5   & 1016.0  & 1.76   & 0.0    & 0.9   & Astrometry   \\
		KIC 7917485 b    & 11.8  & 840.0   & 2.03   & 0.15   & 1.63  &  Pulsation phase modulation \\
		Kepler-419 c     & 7.3   & 675.47  & 1.68   & 0.184  & 1.22  & TTV          \\
		Kepler-46 c      & 0.376 & 57.004  & 0.2799 & 0.0145 & 0.902 & TTV          \\
		OGLE-2006-109L c & 0.271 & 4931.0  & 4.5    & 0.15   & 0.51  & Microlensing \\
		$\beta$ Pic b    & 7.0   & 13288.0 & 13.18  & 0.323  & 1.73  & Imaging      \\
		\hline
	\end{tabular}
\end{table*}

\subsection{Prior distribution of \texorpdfstring{$q$}{q} / mass of planet 2}
\label{sec:methods:prior_q}

\begin{figure}
	\includegraphics[width=\columnwidth]{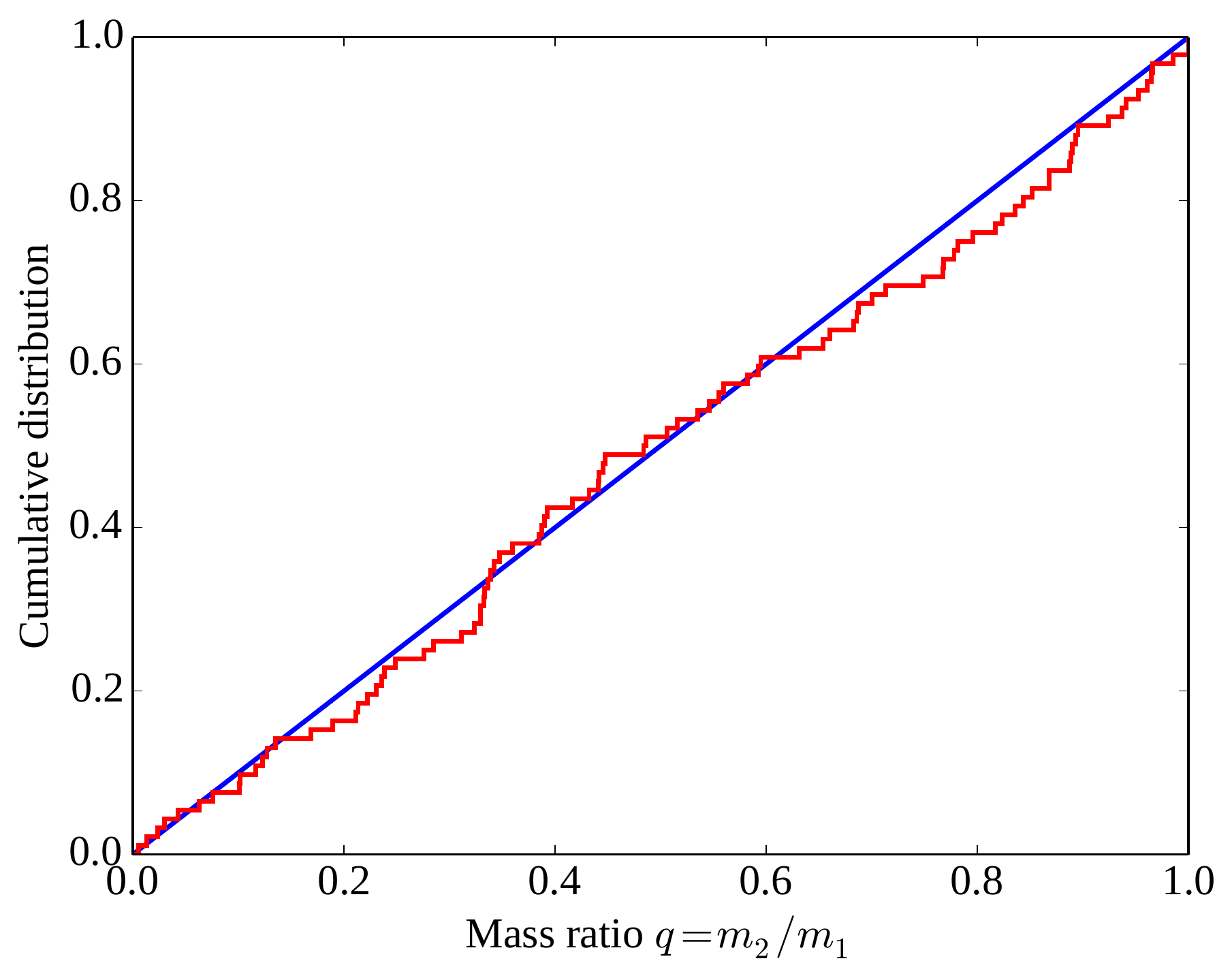}
    \caption{Distribution of mass ratios between neighbouring giant planets (red). Table \ref{tab:dataset} has more information about our data set, and section \ref{sec:methods:data} describes our data selection. The dataset includes 171 planets in multiple-giant systems, and 92 planet pairs. The mass ratio is defined as $q = m_2 / m_1$ where $m_2 < m_1$. For comparison, the uniform distribution is shown in blue.}
    \label{fig:ratios}
\end{figure}

Our goal is to use an informed prior $P(q)$. Our dataset includes 171 planets with giant planet companions. Figure \ref{fig:ratios} shows the cumulative distribution of the mass ratios of neighbouring planets in our dataset. The Kolmogorov-Smirnov test cannot distinguish this distribution from the uniform distribution (p-value of 0.96). Therefore, a uniform Bayesian prior is a reasonable choice. We choose 10 values for $q$ spread uniformly between 0 and 1,

\begin{equation}
	q \in \{0.1, 0.2, 0.3, \cdots, 1.0 \} \;\;\;\;\;{\rm and}\;\; m_2 = q m_1
\end{equation}

\subsection{Mass and semimajor axis of planet 1}
\label{sec:methods:p1}

The collision cross section between two planets is

\begin{equation}
	\sigma = \pi R^2 \left(1 + \frac{v_{\rm esc}^2}{v_\infty^2} \right)
           = \pi R^2 \left(1 + \Theta \right),
\end{equation}
where $R$ is the radius of the planet, $v_{\rm esc}$ is the escape speed at the planet surface, and $v_\infty$ is the relative speed of the two planets before gravitational focusing. The focusing factor $\Theta = v_{\rm esc}^2 / v_\infty^2$ is commonly known as the Safronov number. In the case of a dynamical instability, where planets approach each other at speeds comparable to the orbital speed, the Safronov number becomes

\begin{equation}\label{eqn:Theta}
	\Theta \sim \frac{1}{2} \left( \frac{m}{\Ms} \right) \left( \frac{a}{R} \right).
\end{equation}

When $\Theta \ll 1$, close encounters between planets are are likely to lead to collisions, and when $\Theta \gg 1$ ejections are more common \citep[e.g.][]{Safronov_1972,Binney_2008,Ford_2008}. The difference is important because collisions are more likely to leave planets in less eccentric orbits than ejections, since they conserve the total angular momentum in the system while ejected planets carry angular  \citep[e.g.][]{Ford_2008,Matsumura_2013}. We would like our simulations to have planet masses (in terms of $\mu$) and Safronov numbers comparable to the observed population, so that our runs have the correct proportion of collision and ejection events. Unfortunately, most planets in our sample do not have measured radii. Therefore, we make the simplifying assumption that most planets have a density similar to Jupiter, and introduce the quantity

\begin{equation}\label{eqn:theta}
    \theta = \left( \frac{m}{\Mj} \right)^{2/3} \left( \frac{\Ms}{\Mo} \right)^{-1}
    		\left( \frac{a}{\AU} \right) 
\end{equation}

With this definition, a Jupiter-mass planet orbiting a Sun-like star at 1 AU has $\theta = 1$. As a point of reference, table \ref{tab:theta} compares $\Theta$ and $\theta$ for the giant planets in the solar system. The two quantities are broadly comparable, but $\theta$ can be calculated for the planets in our sample.

\begin{table}
	\centering
	\caption{Comparison of $\Theta$ and $\theta$ (equations \ref{eqn:Theta} and \ref{eqn:theta}) for the giant planets in the solar system. The advantage of $\theta$ is that it can be calculated for exoplanets without measured radii.}
	\label{tab:theta}
	\begin{tabular}{lll}
		\hline
        Planet & $\Theta$ & $\theta$ \\
		\hline
        Jupiter & 5.37 & 5.20 \\
        Saturn  & 3.57 & 4.29 \\
        Uranus  & 2.48 & 2.45 \\
        Neptune & 4.72 & 4.28 \\
		\hline
	\end{tabular}
\end{table}

\begin{figure*}
	\includegraphics[width=\columnwidth]{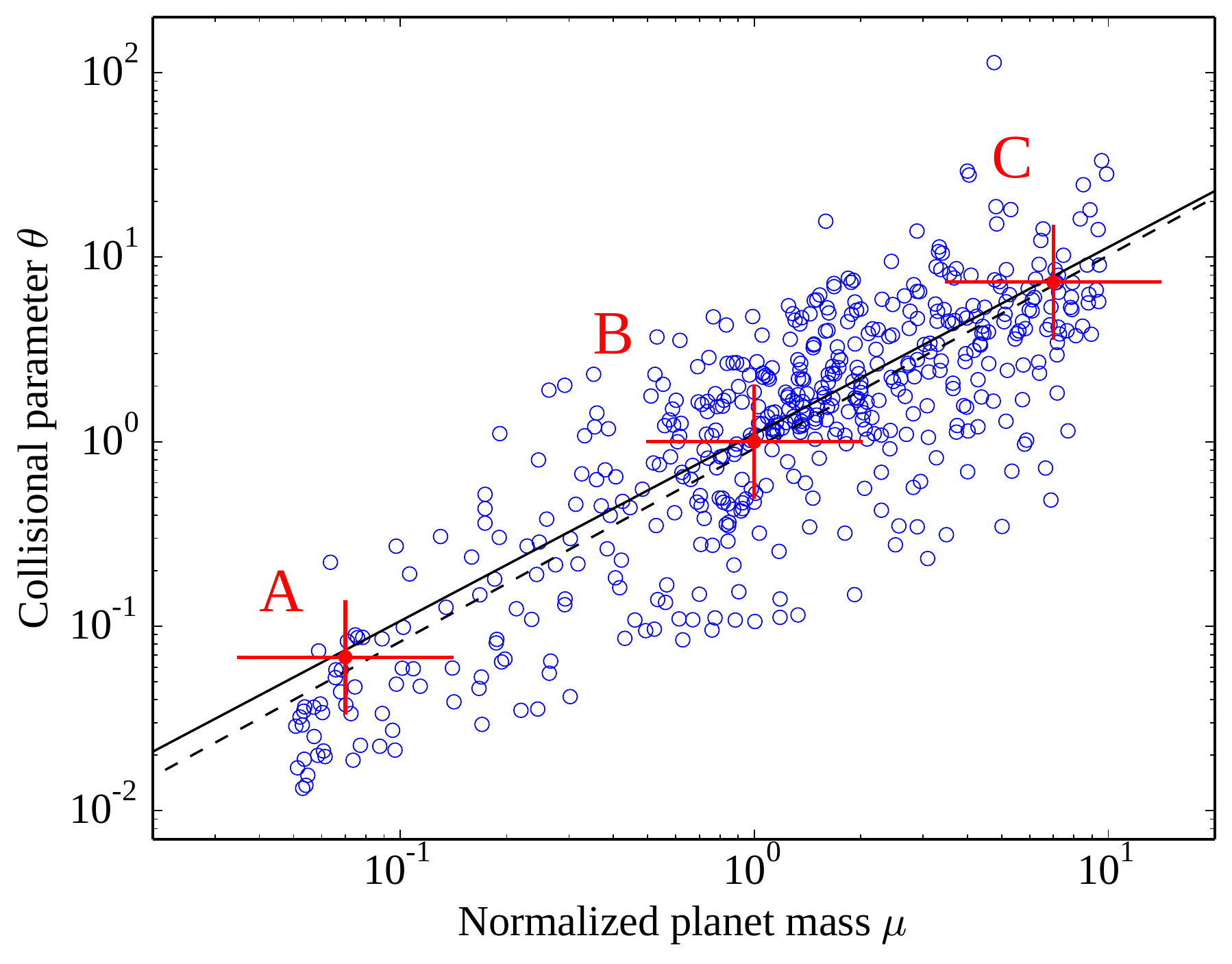}
	\includegraphics[width=\columnwidth]{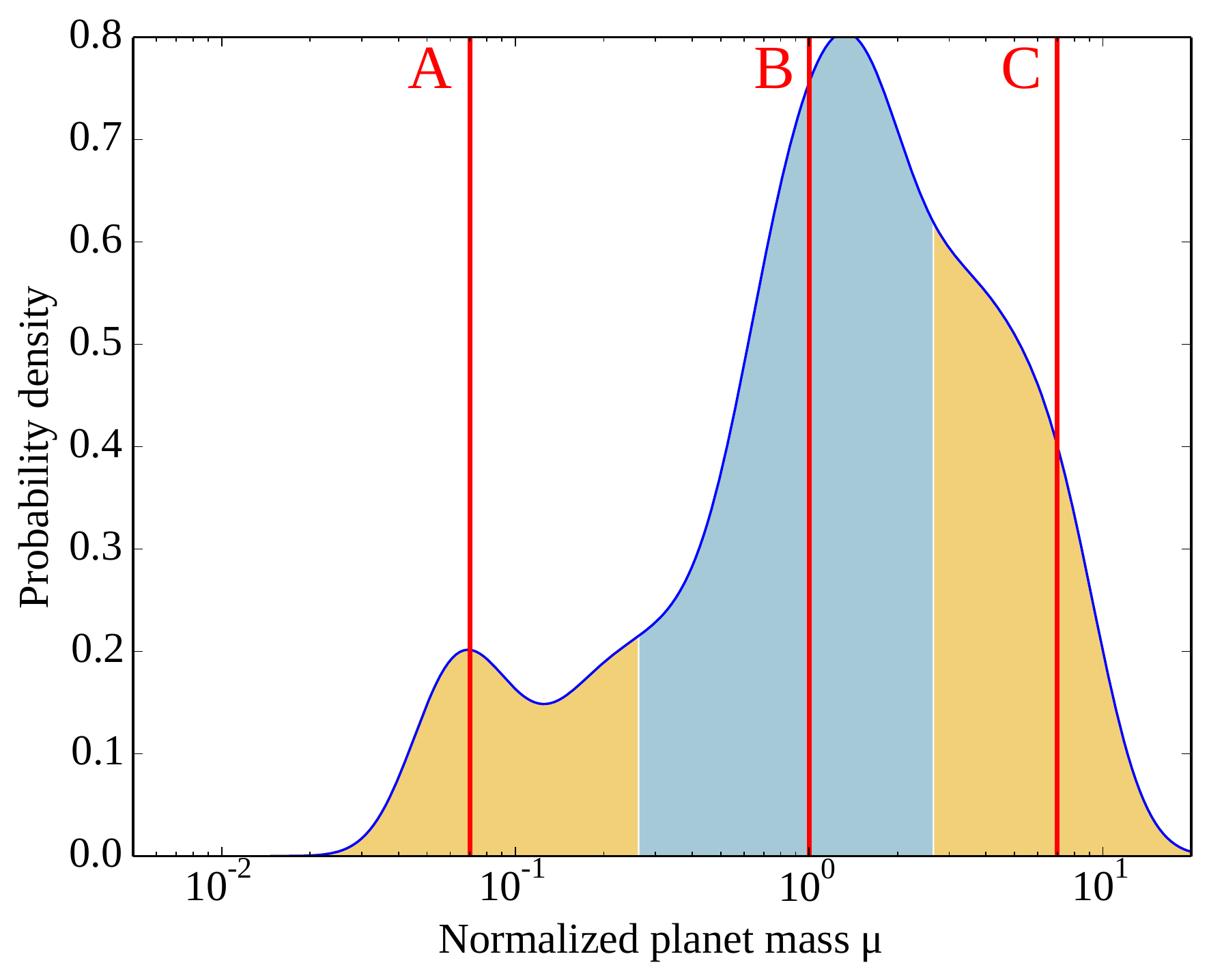}
    \caption{\textit{Left}: Distribution of $\mu$ and $\theta$ (Equations (\ref{eqn:mu}) and (\ref{eqn:theta})) for all observed giant planets ($0.05 < \mu < 10$) with determined masses excluding hot Jupiters (period $\le$ 10d), marked as blue open circles. The solid and dashed black lines are the best-fit lines using the Theil-Sen estimator and ordinary least squares respectively. The red crosses mark the ($\mu$,$\theta$) for our three sets of runs (Table \ref{tab:sets}). We chose these values to strike a balance between covering the full range of the observed ($\mu$, $\theta$) and concentrating our runs in the most densely populated parts of the parameter space. \textit{Right:} Probability density function of $\mu$ (using kernel density estimation). The vertical red lines mark the $\mu$ values for our runs (table \ref{tab:sets}). The three coloured regions mark the points that are closest to sets A, B, and C on a log scale. These regions contain 14.6\%, 56.1\%, and 29.3\% of the giant exoplanets respectively. For this reason, most of our runs are in set B.}
    \label{fig:obs}
\end{figure*}

Figure \ref{fig:obs} shows the distribution of $\mu$ and $\theta$ for the exoplanets in our dataset. Clearly there is a wide range of values for both $\mu$ and $\theta$. For this reason, we decided to do three different sets of runs (calling the sets A, B, and C) with the mass and semimajor axis of planet 1 spread along the best-fit line on figure \ref{fig:obs}. Planet 1 is the more massive planet in each run, and is the one more likely to survive. Our choices for $m_1$, and $a_1$ --- shown in table \ref{tab:sets} --- are a compromise between covering the full range of observed $\mu$ and $\theta$, while also having our runs concentrated in the regions of the parameter space that have more planets. As it turns out, 56\% of the planets in our sample are closer to set B (on a log scale). Therefore, we perform most of our runs in set B (see table \ref{tab:sets}).

\begin{table*}
	\centering
    \caption{We divide the exoplanet population into three sets (A, B, C) based on the normalized mass $\mu$ Eqn.\,\ref{eqn:mu} (see figure \ref{fig:obs}). For each set we conduct N-body simulations with a planet mass and semimajor axis ($m_1$, $a_1$) representative of that set -- the stellar mass is always $1 M_\odot$. We conduct at least 1,000 simulations per set and select the runs that had an instability after 0.5 Myr, and made sure that there are more simulations than observed exoplanets.}
	\begin{tabular}{ccc|ccc}
		\hline
        & \multicolumn{2}{c}{Exoplanet observations} &
        \multicolumn{3}{c}{Simulations} \\
         Set & Mass range & Exoplanets & $m_1 / \Mj$ & $a_1 / \AU$ & Usable runs \\
		\hline
         A   & $\mu < 0.265$          &  55 & 0.07 &  0.4 & 274 \\
         B   & $0.265 \le \mu < 2.65$ & 241 & 1    &  1   & 410 \\
         C   & $\mu \ge 2.65$         & 131 & 7    &  2   & 286 \\
		\hline
	\end{tabular}
	\label{tab:sets}
\end{table*}

\subsection{Other orbital elements}
\label{sec:methods:p2}

We have already covered how we select $m_1$, $a_1$, and $m_2$. The remaining orbital elements are chosen to make our simulations consistent with planet formation. Giant planets form inside a protoplanetary disk, with typical lifetimes in the order of $\sim$ 3-6 Myr \citep[e.g.][]{Hartmann_1998,Haisch_2001,Mamajek_2009}. Orbital configurations that lead to orbit crossing on a timescale significantly shorter than 3 Myr would lead to ejections and collisions during the disk phase, with enough time for the disk to subsequently dampen the eccentricity of the remaining planet. For this reason, we require that the system experience no collisions or ejections for at least 0.5 Myr.

\begin{figure}
	\includegraphics[width=\columnwidth]{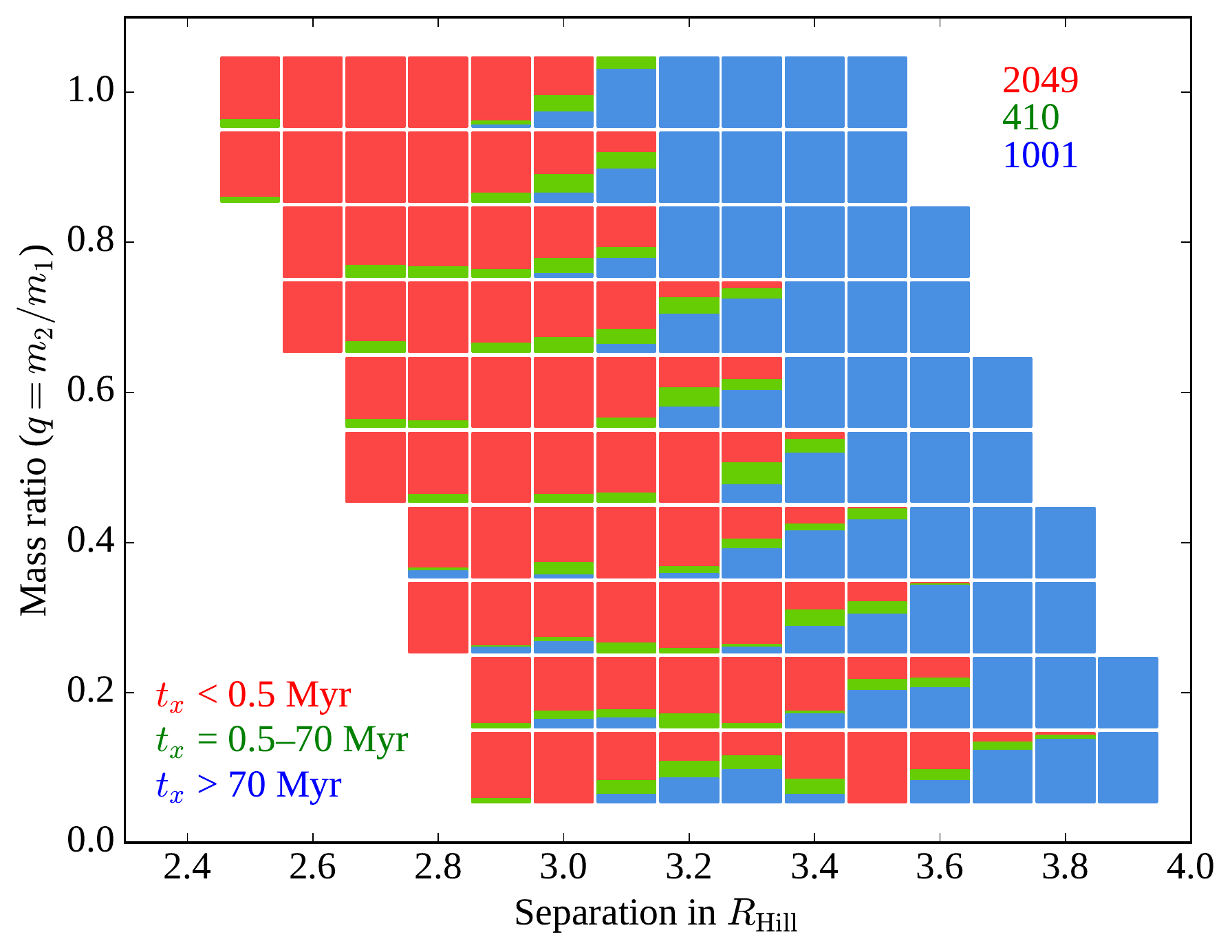}
    \caption{This figure shows the range of mass ratios ($q = m_2/m_1$) and orbital separations that we used for the N-body simulations in set B (our sets are summarized in Table \ref{tab:sets}). We ran simulations for ten mass ratios: $q \in \{0.1, 0.2, \cdots, 1.0\}$. For each $q$ we probed a range of separations in $\Delta$ with the goal of finding configurations that became unstable, but only after 0.5 Myr. Lower mass ratios generally require larger $\Delta$ to survive 0.5 Myr. The coloured boxes mark the $(q,\Delta)$ values where we conducted our simulations. We began with 10 simulations for each value $(q,\Delta)$. Each coloured box is a bar plot showing the fraction of simulations that became unstable in less than 0.5 Myr (red), became unstable after 0.5 Myr (green), or did not become unstable by the end of the 70 Myr run (blue). Only the ``green runs'' are used in our investigation. After the fist 1,100 runs were complete, we chose all the $(q,\Delta)$ that were neither all red, nor all blue, and conducted  1,180 additional runs. The values in the top-right corner show the number of runs in red, green, and blue.}
    \label{fig:set-B}
\end{figure}

To conduct our simulations, we select a range of semimajor axes for planet 2 near the Hill stability limit. The exact range of $a_2$ varies depending on the stability criterion. Figure \ref{fig:set-B} shows the values of $\Delta$ (equation (\ref{eqn:Delta})) for set B. The $\Delta$ values are equally spaced, in steps of 0.1. For each $\Delta$, we solve for $a_2$ and we run ten N-body simulations using the hybrid integrator in {\sc mercury} \citep{Chambers_1999}. Both planets are initially in circular orbits, but we give them mutual inclinations of a few degrees (similar to the solar system). We assign each planet a random inclination $I \in [0^\circ, 5^\circ]$, random longitude of ascending node $\Omega \in [0^\circ, 360^\circ]$, and random mean longitude $\lambda \in [0^\circ, 360^\circ]$. Mutual inclinations are important because an overly flat system will have an unrealistic number of collisions.

The simulations initially run for 10 Myr. If there is an ejection or collision in less than 0.5 Myr, we do not use the run. If there are no collisions or ejections, we extend the runs in 20 Myr increments up to 70 Myr. Figure \ref{fig:set-B} shows a bar plot for each value of $(\Delta, m_2)$ that indicates the fraction of runs that became unstable in less than 0.5 Myr (red), or 0.5-70 Myr (green), or had no instability at the end of the 70 Myr run. Notice that smaller $m_2$ requires larger $\Delta$ to be stable for 0.5 Myr. This is the reason why the range of $a_2$ varies with $m_2$. In a similar way, changing $m_1$ also affects the stability region, so the range of $a_2$ that are consistent with planet formation change as well.

%
%
\section{Result of planet-planet scattering}
\label{sec:results}

In this section we present the results of our N-body simulations. In section \ref{sec:results:two-planets} we apply the Bayesian algorithm described in sections \ref{sec:bayesian} and \ref{sec:methods} to our two-planet simulations and we construct an estimate for the exoplanet IMF. Then in section \ref{sec:results:three-planets} we present a more limited set of simulations with three giant planets and discuss how a three-planet instability differs from the two-planet case.

\subsection{Two-planet systems}
\label{sec:results:two-planets}

\begin{figure}
	\centering
	\includegraphics[width=\columnwidth]{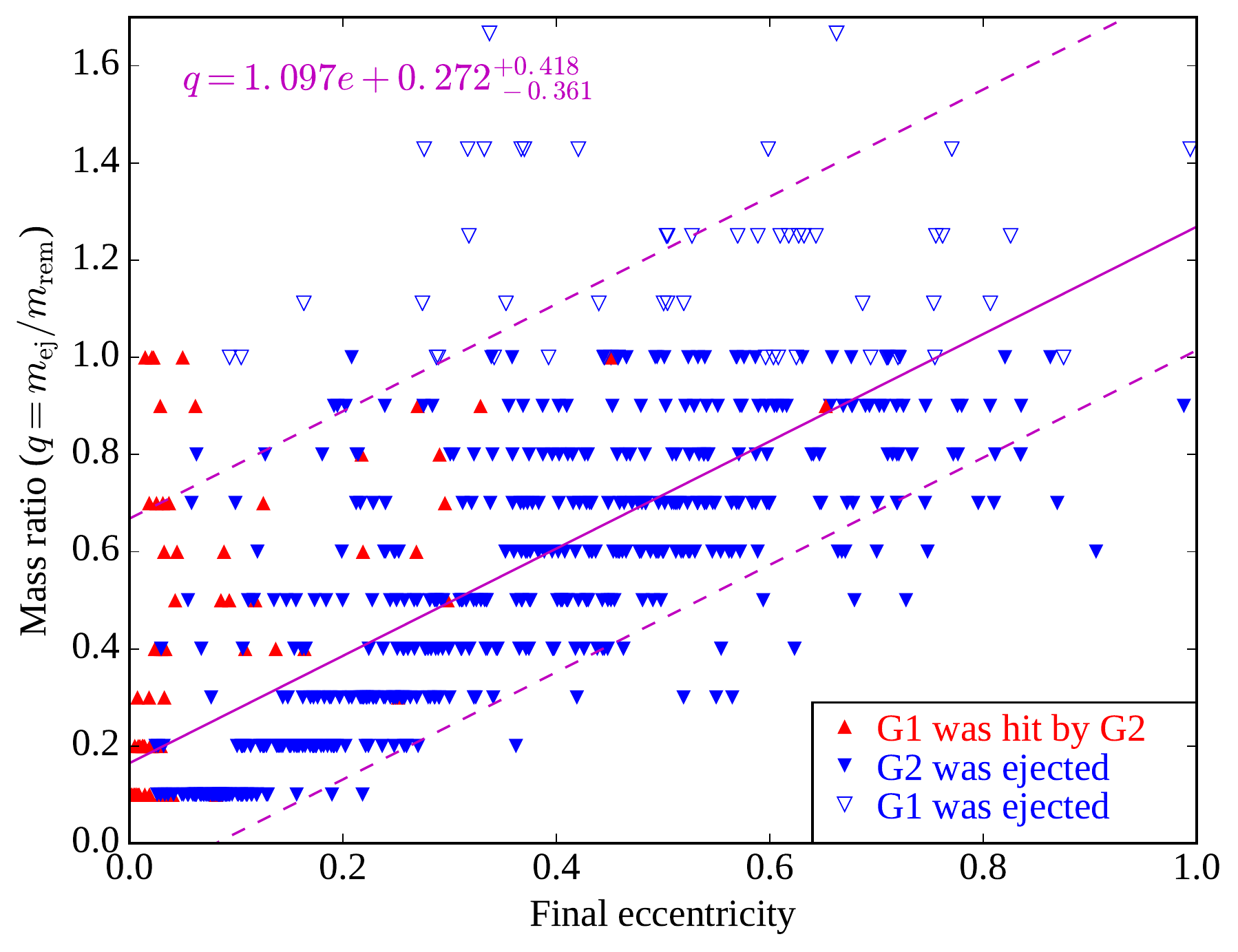}
    \caption{Final outcome of the simulations in set B (see Table \ref{tab:sets}) where an instability occurred after 0.5 Myr. The inner planet (G1) has a mass of 1 $\Mj$ and the second planet (G2) has a mass between 0.1 and 1 $\Mj$. The ratio $q = m_{\rm ej} / m_{\rm rem}$ is the mass of the ejected planet over the mass of the planet that remains, which in most cases is the outer planet. In the case of a collision, $q = m_2 / m_1$ is the mass of the outer planet over that of the inner. The horizontal axis is the final eccentricity of the planet that remains. The plot shows a clear trend where larger mass ratios lead to more eccentric survivors. The solid magenta line is the Theil-Sen fit, and the dashed lines contain the middle 90\% of the points.}
    \label{fig:trend}
\end{figure}

Here we present our main results. Figure \ref{fig:trend} shows the final eccentricity of the surviving planet for every single run in set B that had an instability after 0.5 Myr. By far the most common result was for the lower mass planet to become ejected from the system. In some cases the two planets collided, and in others it was the inner, more massive planet that was ejected. As a general rule, collisions are associated with low-eccentricity outcomes, even for equal-mass planets. The connection between collisions and dynamically quiet histories has already been noted by other authors in the context of three-planet instabilities \citep{Matsumura_2013,Carrera_2016}.

Perhaps the most important feature of Figure \ref{fig:trend} is the correlation between the eccentricity of the remaining planet $e$ and the mass ratio of the planet that was ejected or destroyed over the mass of the surviving planet. In most cases $q = m_2 / m_1$ where planet 2 is the outer planet; but in the runs where the inner (more massive) planet is ejected, we write $q = m_1 / m_2$. The correlation between $e$ and $q$ is not surprising, but it is important --- it is what allows us to produce a probability distribution of $q$.

\pagebreak

\begin{figure}
	\centering
	\includegraphics[width=\columnwidth]{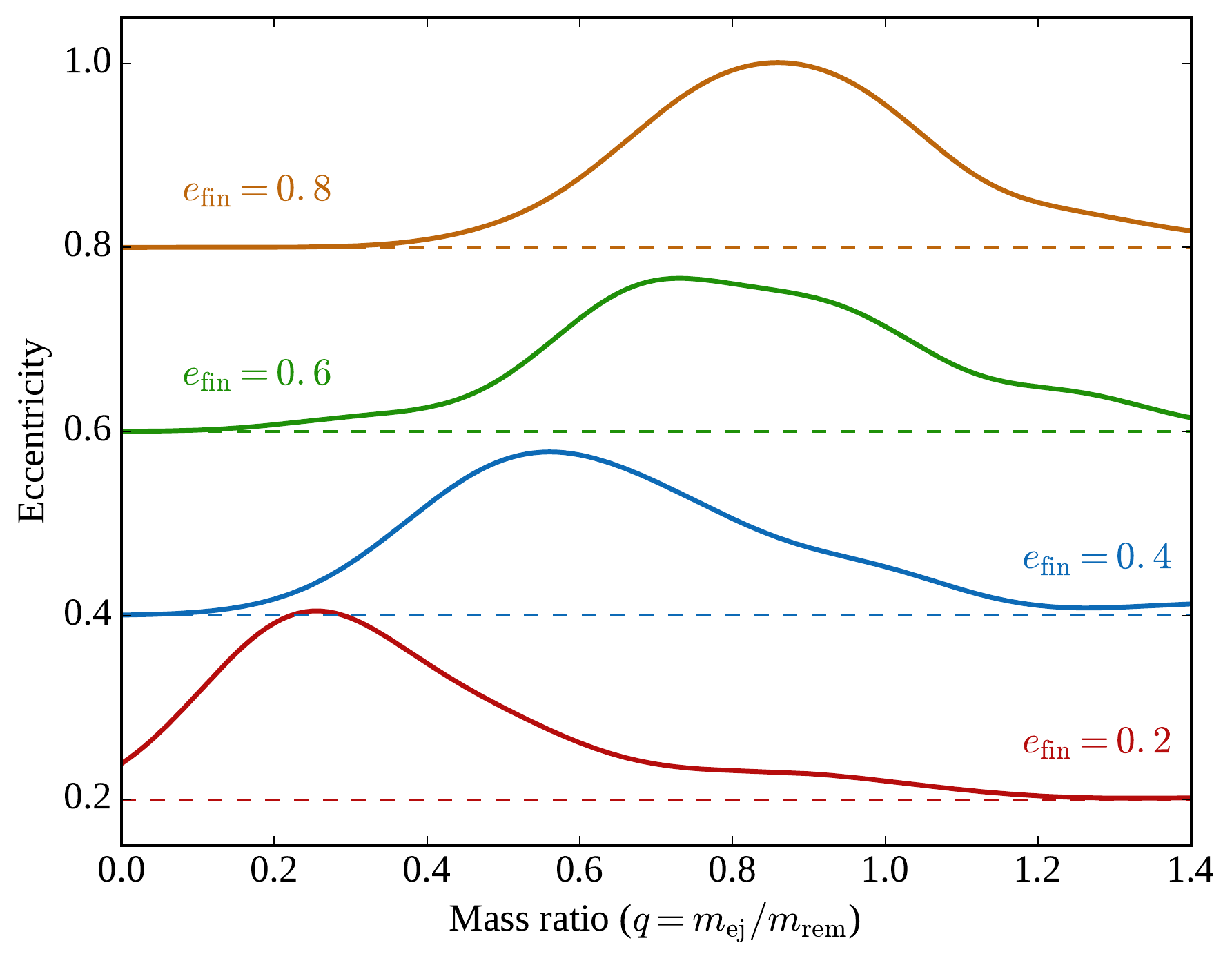}
    \caption{Probability distribution of $q = m_{\rm ej} / m_{\rm rem}$ (mass ratio of the ejected planet and the remaining planet), as a function of the final eccentricity $e_{\rm fin}$ of the remaining planet. The curves are kernel density estimates (KDE) derived from the simulations in set B. For each value of $e_{\rm fin}$ we took all the data points from Figure \ref{fig:trend} that fall within $e_{\rm fin} \pm 0.1$. The KDEs have a fixed bandwidth of $h = 0.1$. To improve clarity, we chose to stack the plots vertically: for each value of $e_{\rm fin}$ we plot $y = e_{\rm fin} + k(q)/10$ where $k()$ is the probability density. The general trend is that planets that have higher eccentricities point to a larger mass for the ejected planet. In particular, given $e_{\rm fin}$ we can estimate $q$. Applying this procedure to every giant exoplanet with a measured mass yields an estimate of the giant planet IMF.}
    \label{fig:ratio-kde}
\end{figure}

Figure \ref{fig:ratio-kde} makes this point more concrete. Here we choose four sample eccentricities ($e_{\rm fin} \in \{0.2, 0.4, 0.6, 0.8\}$), and for each one we compute a probability distribution $k(q)$ using the data in set B (Figure \ref{fig:trend}). Conceptually, for any given value of $e$ one can draw a vertical line in Figure \ref{fig:trend} and select all the points near that line. We use kernel density estimation (KDE) to convert these points into a smooth distribution. A KDE is conceptually equivalent to replacing each point with a Gaussian distribution with standard deviation $h$ (called the bandwidth). In Figure \ref{fig:ratio-kde} we chose a wide bin ($e_{\rm fin} \pm 0.1$) to focus on the overall shape of the curve, but notice that in Figure \ref{fig:trend} there is a relatively small number of points for any given $e_{\rm fin}$. As the number of simulations continues to increase, our statistics will gradually improve. Finally, to produce the exoplanet planet IMF, we generalize the procedure:

\begin{itemize}
\item For planets with $\mu \ge 0.05$ and $e \ge 0.05$ we randomly select $N = 1000$ runs (see section \ref{sec:bayesian}) with $|e_{\rm sim} - e_{\rm obs}| < 0.02$ and compute the initial planet masses $m_1$ and $m_2$.

\item For planets with with $\mu < 0.05$ or $e < 0.05$ we assume that there was no planet-planet scatter, and we simply copy the planet mass $N = 1000$ times.
\end{itemize}

\begin{figure*}
	\centering
	\includegraphics[width=\columnwidth]{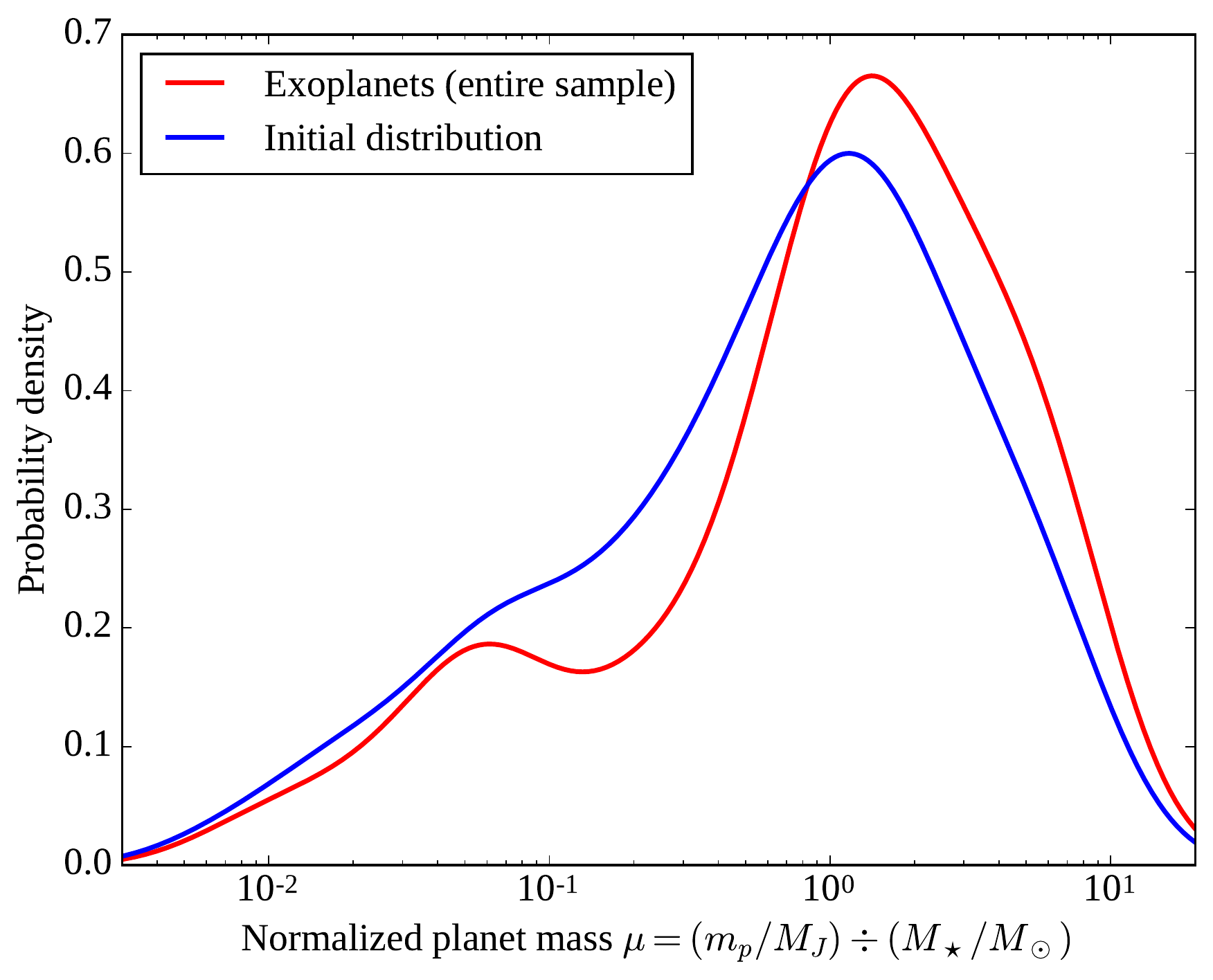}
	\includegraphics[width=\columnwidth]{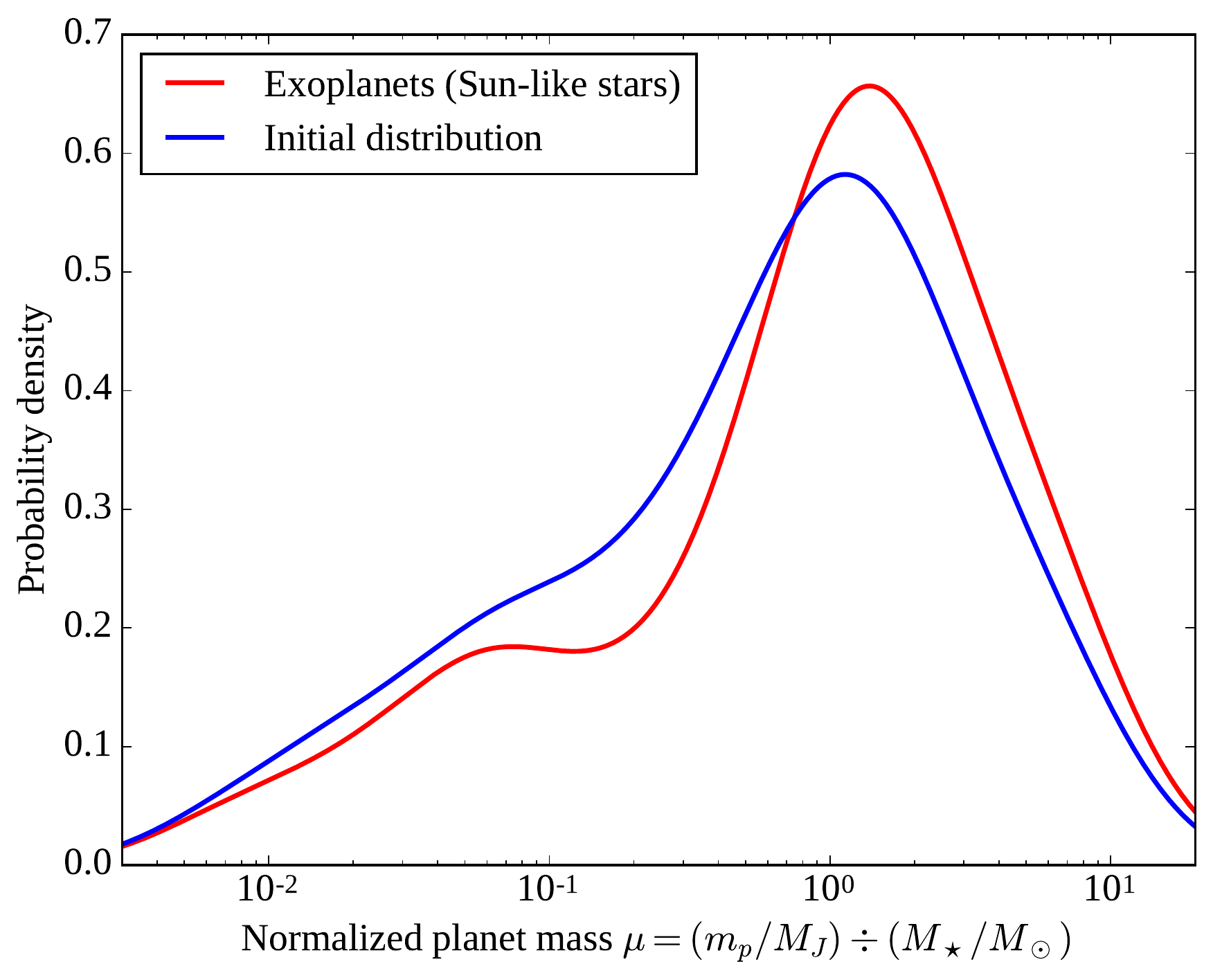}
    \caption{Observed distribution of exoplanet masses as seen today (red) and the estimated initial distribution (blue) for all stellar types (left) and for stars between 0.9 and 1.1 $\Mo$ (right). We measure the planet mass relative to the stellar mass (see axis label). Here we assume that only planets with $e \ge 0.05$ and $\mu \ge 0.05$ (Neptune-like) acquired their eccentricity through a collision or scattering with another planet. Table \ref{tab:dataset} has more information about our data set, and section \ref{sec:methods:data} describes our data selection. For every planet we obtain the probability distribution of $q = m_{\rm ej} / m_{\rm rem}$ as illustrated in Figure \ref{fig:ratio-kde}. Here we use eccentricity bins $e_{\rm fin} \pm 0.02$ and bandwidth $h = 0.05$.}
    \label{fig:IMF}
\end{figure*}

Together, these sets give a synthetic population that approximately follows the initial mass function of the exoplanets in our dataset. As in Figure \ref{fig:ratio-kde}, we compute the kernel density $k(\mu)$. Figure \ref{fig:IMF} shows the final result for all the planets in our dataset, and for the planets orbiting Sun-like stars. One interesting feature of these plots is that the observed exoplanet population appears to have a ``valley'' at around $\mu = 0.2$ (sub-Saturn planets), but the synthetic IMF suggests that this valley is not primordial, but was carved out later by planet-planet scatterings. If this interpretation is correct, there should be a population of free-floating Saturn-like planets that were ejected from their host system by a more massive planet. However, note that \citet{Veras_2012} have shown that dynamical instabilities cannot be the primary source of the observed population of free-floating giant planets \citep{Clanton_2017}, as that would require an implausible number of planetary ejections per system.

\subsection{Three-planet systems}
\label{sec:results:three-planets}

Our results so far only apply to two-planet systems, and it is important to understand to what extent they generalize to multiple-planet systems. In this section we present a set of N-body simulations with three giant planets, and we discuss the similarities and differences between two-planet and three-planet systems.

All our three-planet simulations have a Jupiter-mass planet at 1 au, and two exterior planets with half the mass of Jupiter. We call this set \texttt{J+J/2+J/2}. The planets have a uniform separation in terms of their mutual Hill radii (equal $\Delta$, see Equation (\ref{eqn:Delta})). We compare these runs against the two-planet runs with $q = 1$ (we call them \texttt{J+J}) and $q = 0.5$ (we call them \texttt{J+J/2}). The three-planet runs are generally less stable, and we need to be more widely spaced ($\Delta > 4$) for them to last 0.5 Myr. That said, in terms of orbital energy and angular momentum the difference is not very large. For example, in a \texttt{J+J/2+J/2} system with $\Delta = 4.1$ the two outer planets have 79\% of the binding energy and 113\% of the angular momentum of the outer planet in a \texttt{J+J} system with $\Delta = 2.5$.

\begin{figure}
	\centering
	\includegraphics[width=\columnwidth]{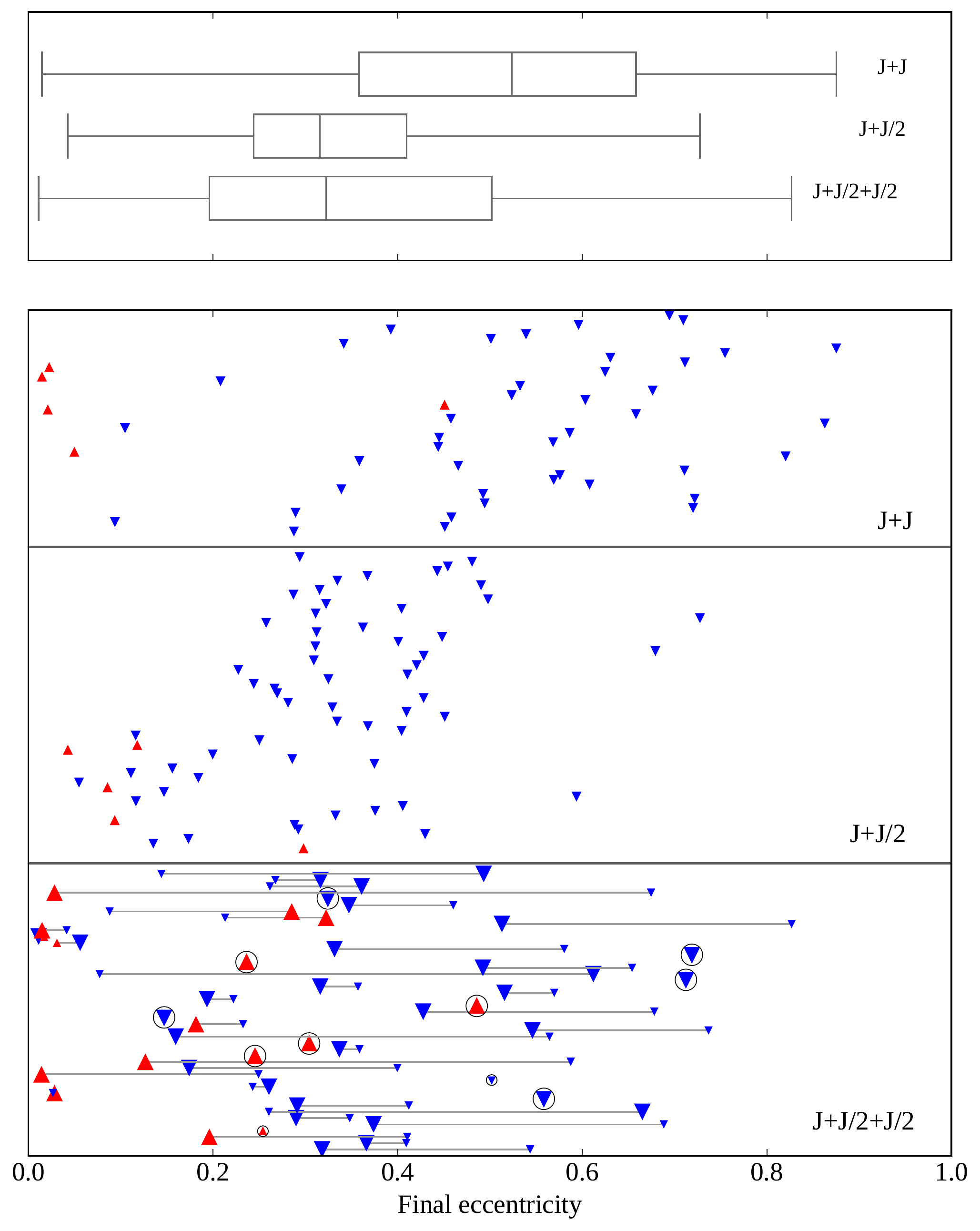}
    \caption{Final eccentricity of the surviving planets in three-planet and two-planet systems, for systems that went unstable after 0.5 Myr. \texttt{J+J/2+J/2} denotes runs with one Jupiter-mass planet and two half-Jupiter planets, \texttt{J+J/2} denotes the runs with one Jupiter and one half-Jupiter planet, and \texttt{J+J} denotes runs with two Jupiter-mass planets. The top plot summarizes the eccentricities as a box plot, showing the four quartiles. The bottom plot shows the final eccentricity of every surviving planet. The symbol denotes whether the planet experienced a collision (up-triangle) or not (down-triangle). For \texttt{J+J/2+J/2}, most runs had two surviving planets; these are marked with a grey line connecting the planets and the Jupiter-mass planet is marked with a larger symbol.}
    \label{fig:3p-ecc}
\end{figure}

Figure \ref{fig:3p-ecc} shows the final eccentricity of the surviving planets in each set of runs. Broadly speaking, \texttt{J+J/2+J/2} produces eccentricities similar to those in \texttt{J+J/2}, and somewhat lower than those in \texttt{J+J}. In \texttt{J+J/2+J/2} there are a few runs where only one planet survived. These planets are more eccentric, and are generally consistent with \texttt{J+J}.

%
%
\section{Some caveats}
\label{sec:discussion}

In this section we discuss sources of bias or uncertainty that might affect the accuracy of our results. When possible, we discuss ways that these sources of error might be corrected or at least measured.

\subsection{Observational errors and biases}
\label{sec:discussion:obs}

The exoplanet sample used in this investigation is plagued with observational biases. There are completeness issues because massive planets are easier to detect than lower-mass planets. In addition, eccentricities from radial velocity surveys are unreliable and the RV signal caused by two planets in circular orbits can be difficult to distinguish from the signal caused by a single planet in an eccentric orbit. De-biasing RV surveys is a difficult problem that lies beyond the scope of the present investigation.

We did explore the completeness issue with a reduced subset of observed planets. We selected planets in the range of periods 10 to 300 days and masses between 0.1 and 10 $\Mj$, and repeated our procedure. This subset suffers from small number statistics, as there are only 27 planets in the set; nonetheless, our key conclusion continues to hold for this data set --- that the observed exoplanet population has a deficiency in sub-Saturn-mass planets, and that that this gap is not primordial, but is the result of ejections.

\subsection{System architecture}
\label{sec:discussion:arch}

This investigation is limited to planetary systems with two dynamically dominant planets. In section \ref{sec:results:three-planets} we quantified some of the key differences between 2-planet and 3-planet systems. That said, a full investigation is not feasible because the problem is too degenerate when there are more than two giant planets in the system.

\begin{figure}
	\centering
	\includegraphics[width=\columnwidth]{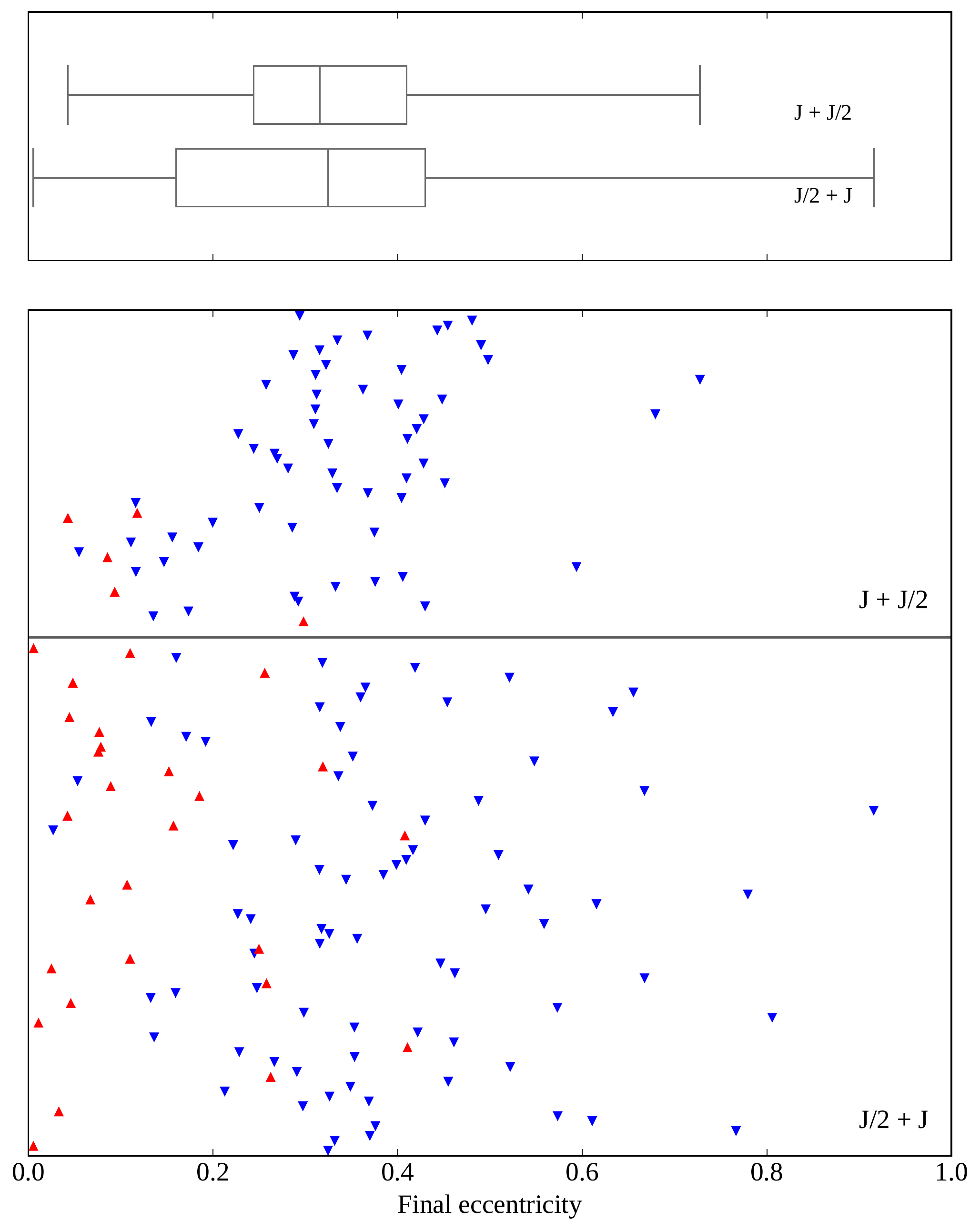}
    \caption{Final eccentricity of the surviving planets. Runs \texttt{J + J/2} have a Jupiter-mass planet at 1 AU and a half-Jupiter on an exterior orbit, 2.7-3.7 mutual Hill radii apart. Runs \texttt{J/2 + J} have the planet orbits reversed, so that the inner planet is the more massive. The top plot summarizes the eccentricities as a box plot, showing the four quartiles. The bottom plot shows the final eccentricity of every surviving planet. A red up-triangle denotes a collision and a blue down-triangle denotes an ejection. The ejected planet is always the less massive one.}
    \label{fig:qs-ecc}
\end{figure}

In all our simulations the inner planet is the most massive. It is possible that if the order of the planets was reversed, the simulation results would be different. To test this, we conducted 550 simulations with the same parameters as \texttt{J+J/2} (i.e.\,set B with $q = 0.5$), but we reversed the order of the planets, so that the inner plant has a mass of 0.5 $\Mj$ and the outer planet has 1 $\Mj$. We call this new set \texttt{J/2+J}. The two sets of runs give similar results. The eccentricity distributions are similar, and in both cases it is always the less massive planet that becomes ejected. One difference is that \texttt{J/2+J} is more collisional than \texttt{J+J/2}. In \texttt{J+J/2} there are 2.5 ejections per collision, while in \texttt{J/2+J} there are only 1.3 ejections per collision. Figure \ref{fig:qs-ecc} gives an overview of the simulation results. The two sets of runs have the same mean eccentricity, but \texttt{J/2+J} may have a wider eccentricity distribution.

\subsection{Eccentricity damping}
\label{sec:discussion:damp}

The next complication is that gravitational interactions with a planetesimal belt can dampen a planet's eccentricity \citep{Raymond_2010}. In fact, some authors suggest that this might have occurred in the solar system at the time of the Late-Heavy Bombardment \citep{Gomes_2005}. In this way, a planetesimal belt may erase information about the dynamical history of the planetary system. Fortunately, this is mainly a concern for low-mass (sub-Saturn) planets in wide orbits \citep{Raymond_2010}. Hence, it is unlikely that this type of damping has a major impact in our results. Nonetheless, it is desirable to quantify how often this occurs. \citet{Raymond_2010} suggested a search for $K \approx 5 {\rm m\,s}^{-1}$ planets with long periods ($P \sim 10$ yr). Another possibility is to look for planetary systems with highly depleted debris belts, which is an independent indicator of past dynamical instability \citep{Raymond_2011,Raymond_2012} and measure how often these systems have low eccentricities (which could suggest eccentricity damping).

\subsection{Secular effects}
\label{sec:discussion:secular}

Finally, in a multiple planet system, planets can exchange angular momentum over secular timescales. For example, if a planetary system with three giant planets ejects one, the two remaining planets will exchange eccentricity periodically. In other words, some of the eccentricities that we observe today may differ from those at the end of the dynamical instability. To tackle this one could model the secular evolution of the remaining planets and note the range of eccentricities that the planets acquire. This, in turn, gives a range of possible solutions for the ejected planet mass. We will revisit this idea in future work.

%
%
\section{Summary and conclusions}
\label{sec:conclusions}

In this work we developed a novel Bayesian framework, supported by N-body simulations, to estimate the probable masses of planets that have suffered collisions or ejections, using the present-day masses and orbits of the surviving planets. When applied to the entire exoplanet population, this technique yields the exoplanet initial mass function. We then demonstrated the use of this technique using present-day observations and several thousand N-body integrations.

The observed exoplanet population has a paucity of sub-Saturn-mass planets. We find that this gap is not primordial, and is mainly the result of dynamical instabilities where more massive giant planets eject less massive ones. This in turn implies that there is a population of free-floating sub-Saturn planets that might be detectable by the upcoming WFIRST telescope through its micro-lensing survey \citep{Clanton_2017}.

%
%
\section*{Acknowledgements}

The authors are supported by the project grant ``IMPACT'' from the Knut and Alice Wallenberg Foundation (KAW 2014.0017), as well as the Swedish Research Council (grants 2011-3991 and 2014-5775), and the European Research Council Starting Grant 278675-PEBBLE2PLANET. D.C. acknowledges Dimitri Veras for helpful discussions and guidance in calculating the Hill stability limit. Computer simulations were performed using resources provided by the Swedish National Infrastructure for Computing (SNIC) at the Lunarc Center for Scientific and Technical Computing at Lund University. Some simulation hardware was purchased with grants from the Royal Physiographic Society of Lund.




\bibliographystyle{mnras}
\bibliography{bibliography}


\bsp	
\label{lastpage}
\end{document}